\documentclass[fleqn,usenatbib]{mnras}

\usepackage{newtxtext,newtxmath}

\usepackage[T1]{fontenc}
\usepackage{ae,aecompl}

\usepackage{graphicx}	
\usepackage{amsmath}	
\usepackage{amssymb}	

\title[Deep imaging of Stephan's Quintet]{Revisiting Stephan's Quintet with deep optical images}

\author[P. --A. Duc et al.]{
Pierre-Alain Duc,$^{1}$\thanks{E-mail:pierre-alain.duc@astro.unistra.fr}
Jean-Charles Cuillandre $^{2}$
and Florent Renaud $^{3}$
\\
$^{1}$Universit\'e de Strasbourg, CNRS, Observatoire astronomique de Strasbourg, UMR 7550, F-67000 Strasbourg, France \\
$^{2}$IRFU, CEA, Universit\'e Paris-Saclay, Universit\'e Paris Diderot, AIM, Sorbonne Paris Cit\'e, CEA, CNRS, \\
Observatoire de Paris, PSL Research University, F-91191 Gif-sur-Yvette Cedex, France\\
$^{3}$Department of Astronomy and Theoretical Physics, Lund Observatory, Box 43, SE-221 00 Lund, Sweden
}

\date{Accepted for publication in MNRAS letters}

\pubyear{2017}

\begin{document}
\label{firstpage}
\pagerange{\pageref{firstpage}--\pageref{lastpage}}
\maketitle

\begin{abstract}
Stephan's Quintet, a compact group of galaxies, is often used as a  laboratory to study a number of phenomena, including physical processes in the interstellar medium, star formation, galaxy evolution, and the formation of fossil groups. As such, it has been subject to intensive multi-wavelength observation campaigns. Yet, models lack constrains to pin down the role of each galaxy in the assembly of the group. We revisit here this system with multi-band deep optical images obtained with MegaCam on the Canada-France-Hawaii Telescope (CFHT), focusing on the detection of low surface brightness (LSB) structures. They reveal a number of extended LSB features, some new, and some already visible in  published images but not discussed before. An extended diffuse, reddish, lopsided, halo is detected towards the early-type galaxy NGC 7317, the role of which had so far been ignored in models. The presence of this halo made of old stars may indicate that the group formed earlier than previously thought. Finally, a number of additional  diffuse filaments are visible, some close to the foreground galaxy NGC 7331 located in the same  field. Their structure and association with mid--IR emission suggest contamination by emission from Galactic cirrus.
\end{abstract}




\section{Introduction}

Stephan's Quintet (SQ), a compact group of five galaxies located at a distance of 85 Mpc\footnote{We adopt here $H_0=75$ km$\,$s$^{-1}\,$Mpc$^{-1}$. At the assumed distance, 10" correspond to 4.1~kpc}, is arguably the poster child for this class of objects. This system exhibits a remarkable variety of physical processes at all scales, pertaining to cooling and heating of the interstellar medium \cite[e.g.][]{Appleton17}, star formation in shock regions \cite[e.g.][]{Guillard12}, star cluster formation \cite[e.g.][]{Trancho12}, tidal dwarf galaxy (TDG) formation \cite[e.g.][]{Lisenfeld04}, not to mention the repeated galaxy collisions and group formation. Not surprisingly, this remarkable system has been observed at all wavelengths, from the X-rays \cite[e.g.][]{OSullivan09} and UV \cite[e.g.][]{Xu05} to the far--infrared \cite[e.g.][]{Appleton13} and radio H{\sc i} \cite[e.g.][]{Williams02}.

\begin{figure*}
\includegraphics[width=\textwidth]{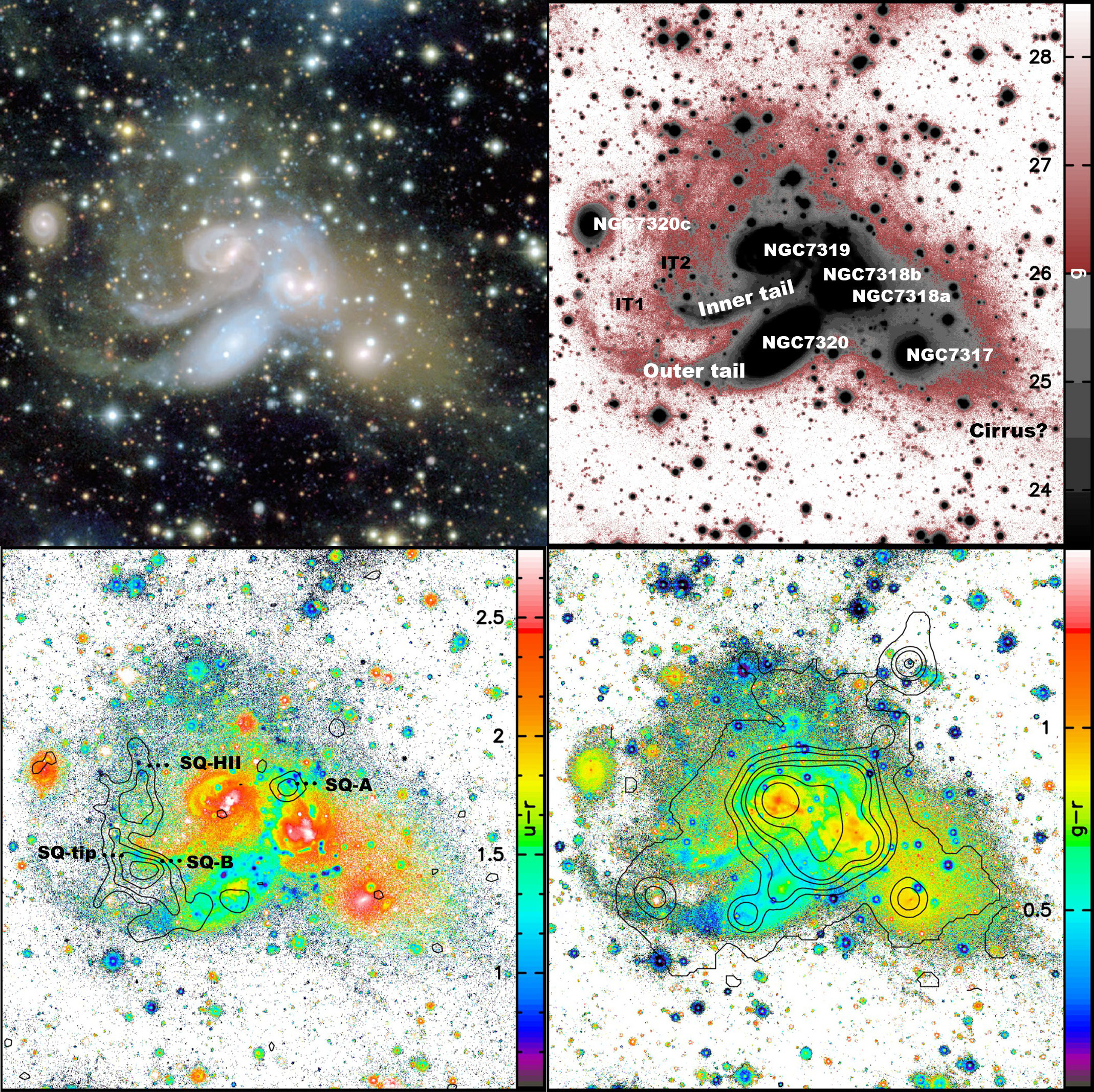}
\caption{{\it Top-left}: composite u+g+r true colour image of the Stephan's Quintet. {\it Top-right}: g-band surface brightness map, with the principle structures labeled. The faintest emission (with surface brightness above 26 mag.arcsec$^{-2}$) revealed by the deep CFHT MegaCam image is shown in red. {\it Bottom-left}: u-r colour map with H{\sc i}/VLA map in the velocity range 6475-6755 km.s$^{-1}$ superimposed. The lowest contour  is 6$\times 10^{19}$ cm$^{-2}$ \citep[adapted from][]{Williams02}. Selected intergalactic star-forming regions are labeled.   {\it Bottom-right}: g-r colour map with archival X-ray/XMM-Newton contours superimposed.  For the surface brightness and colour maps, the scale in mag resp. mag.arcsec$^{-2}$ is indicated to the right. The field of view is 10' $\times$ 10' (250~kpc $\times$ 250~kpc). North is up and East left.}
\label{fig:SQ}
\end{figure*}

Several attempts have been carried out to reproduce numerically the full variety of morphological features of the group, including multiple tidal tails, and complex velocity fields \cite[e.g.][]{Renaud10,Hwang12}, to propose formation scenarios and to explore the underlying physical processes. 
Current models involve a number of events in the last half Gyr, including the collision between the spiral galaxy NGC~7319 and NGC~7320c, creating the so-called outer tail, between NGC~7319 and the early-type galaxy NGC~7318a creating the inner tail, and finally the late high speed collision with the spiral NGC~7318b, producing shocks in the gas stripped by the previous interactions (see Fig.~\ref{fig:SQ}). So far, the lack of observed tidal features between a fifth galaxy, NGC 7317, and the other group members suggested this galaxy had not yet interacted with the group. 

Past studies of SQ have mostly focused on tracers of relatively young events, such as the cold gas (H{\sc i}, molecular), the ionised/shocked  gas or the dust emission. However, few analyses have yet been carried out on the old stellar populations, although they can be used to reconstruct the mass assembly of the group to older times and help constraining the scenarios. Mapping the faint diffuse stellar halos and fine structures around the galaxies is particularly helpful for this purpose. 
Such a study requires images with relatively large fields of view, which was actually the case when, until the 1980s, photography with plates and Schmidt telescopes was available. A rather deep image of the field of SQ was presented by \cite{Arp72}. At that time when the mutual distances between all group member candidates were still discussed, the motivation for such observations was the detection of stellar bridges between the group and the spiral galaxy NGC~7331\footnote{which is now known to be in the foreground and unrelated to SQ, but perhaps not to the other foreground galaxy NGC~7320}. Indeed, \cite{Arp72} detected some streams, but could not firmly conclude on their origin. \cite{Gutierrez02} obtained so far the deepest optical image of SQ, using R band observations with the WFC on the 2.5~m Isaac Newton Telescope at La Palma. With a 2000 second exposure time, they reached a local surface brightness limit of 27.3 mag$\,$arcsec$^{-2}$, allowing them to disclose a diffuse halo around SQ, and to trace back the known tidal tails in the system to much larger distances. The  field of view was however too limited to investigate the structures previously found by \cite{Arp72} between SQ and NGC~7331.

Meanwhile, with the advent of new generation of wide field of view cameras coupled with new techniques optimised for the detection of LSB structures, deep optical imaging of nearby galaxies has been rejuvenated  \citep[e.g.][]{Martinez10,Ferrarese12,Mihos13,Duc15,Trujillo16}. 
We revisit here SQ and its surroundings using multi-band deep optical images obtained  at the CFHT.  We gain new insights, in particular on the role of NGC~7317, and gather clues to prepare for the next generation of models for this group in particular, and of studies of repeated interactions in general.

\begin{figure*}
	\includegraphics[width=0.68\textwidth]{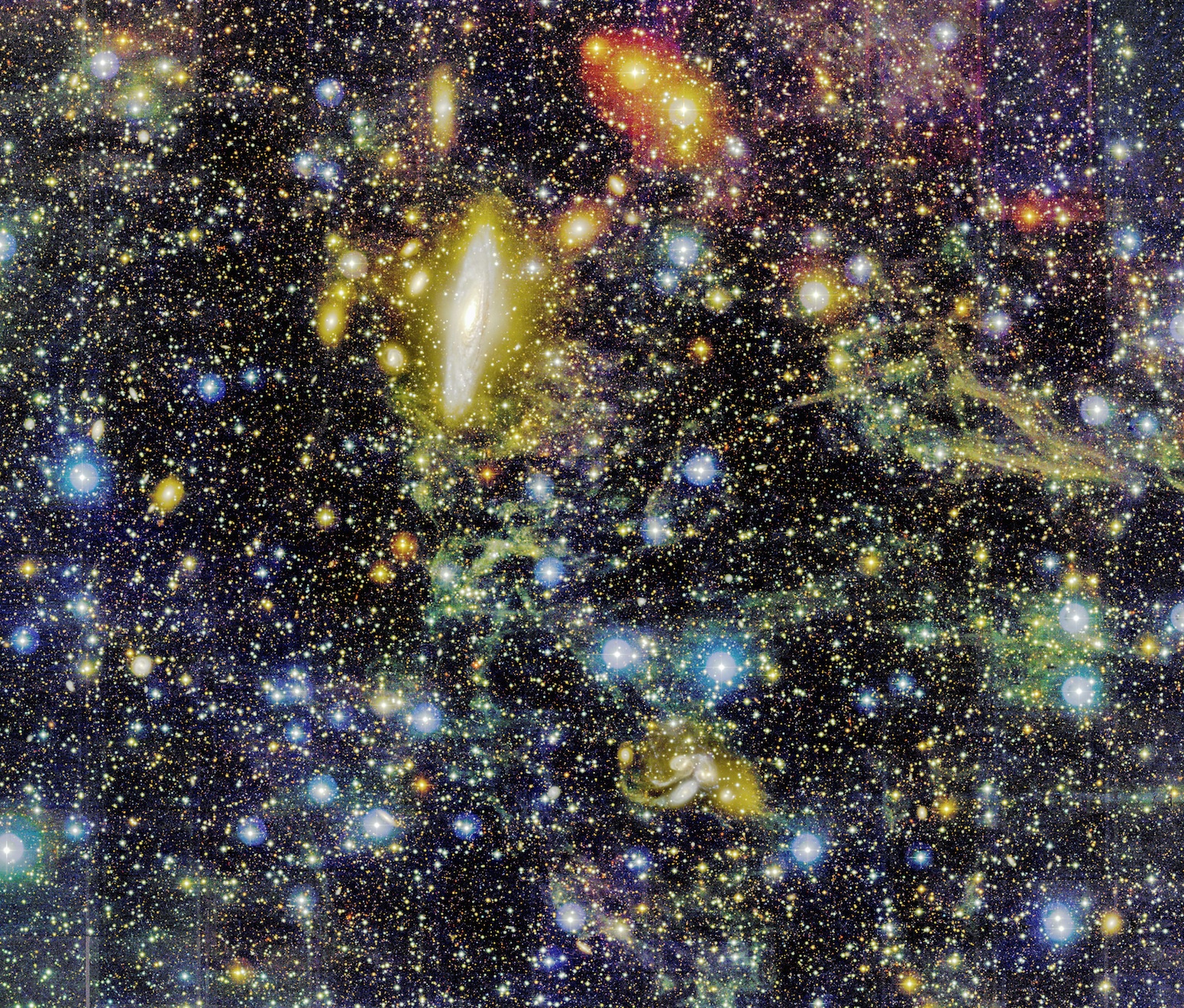}	
  \caption{Composite u+g+r band image of the whole CFHT/MegaCam field, encompassing SQ (bottom, right)  and NGC~7331 (top, left). The field of view is 70' $\times$ 60'. North is up and East left. An arcsinh scaling has been used to enhance the LSB features.}
  \label{fig:field}
\end{figure*}

\section{Observations and data reduction}

Deep multi--band images of the field encompassing SQ and the spiral galaxy NGC~7331 have been obtained with the MegaCam camera installed on the CFHT, through Director Discretionary Time (DDT, PI: J.-C. Cuillandre) from July to December 2015. Images have been acquired with an observing strategy optimised for the detection of LSB objects. 
It relies on (a) acquiring consecutively 7 individual images with  large offsets between them, such that the  distance between each image centre  is at least 10', a scale larger than the  LSB features around SQ studied here (b) building a master sky median-stacking the individual images (c) subtracting a smoothed and rescaled version of the master sky from each individual image before stacking them. 
The procedure,  fully described in \cite{Duc15}, is identical to the one used for the  MATLAS project.
The final stacked images consist of 4 $\times$ 7  frames with total exposure time of 56 min for the u band, and 28 min for the g and r bands.
 Following the DDT policy, observations were obtained with  poor seeing conditions with respect to the  CFHT standards (average seeing 1.8''), but under photometric conditions.
Images were processed by the Elixir-LSB pipeline. The estimated {\it local} limiting surface brightnesses  of  29.0, 28.6 and 27.6 mag$\,$arcsec$^{-2}$ for resp. the u, g and r bands assess the ability of detecting  structures of typical size  5" $\times$ 5".

\section{Results}

The CFHT images provide a panoramic view of  the stellar populations that are present all across the group, up to large distances:  intergalactic young  stars and star clusters, distributed  along blue, narrow and clumpy filaments, old stars present in reddish diffuse tails or extended halos, with a possible contamination by filamentary structures  associated with cirrus emission. 

\subsection{The diffuse halo}

SQ appears to be embedded in a diffuse reddish halo, which is especially prominent towards the South-West, near NGC~7317. This early-type galaxy is off-centred with respect to the outer isophotes of the halo. The diffuse emission extends up to 60~kpc from the galaxy. As seen in the true colour image (Fig.~\ref{fig:SQ}, top-left), but even more clearly on the calibrated u-r and g-r colour maps (Fig.~\ref{fig:SQ}, bottom), the diffuse light is much redder to the South-West than towards the North of the compact group. The average colours (u-r=2.6~mag, g-r=0.9~mag) correspond to typical Gyr old stellar populations. Only a mild negative g-r colour gradient of 0.05 dex per 10 kpc is observed. 

We note  that PSF effects that may affect the light and colour distributions of galaxies \citep[see the study of][based on MegaCam images]{Karabal17} cannot explain the specific properties of the diffuse, off-centred, light around NGC~7317. At its location on the MegaCam image, instrumental ghosts generated by the galaxy nucleus should instead be concentric as illustrated by the  light profiles of the bright  foreground stars surrounding the galaxy.

Since the stellar halo traces a dynamically hot morphology (as opposed to sharp tidal tails), it is likely that its stars have been stripped from the elliptical galaxy NGC~7317 during an interaction with one (or more) of the group members. The lack of tidal arms pointing toward a specific disk galaxy forbids us to designate a culprit though an implication of NGC~7318b should be excluded because of it high radial velocity.  Such arm(s) either never formed (e.g. because of a retrograde encounter), or have been erased by subsequent interactions, which would suggest that NGC~7317 has been involved early in the construction of the group.

\subsection{Young and old tails}

SQ is  known for hosting multiple  intergalactic blue regions, with star-formation triggered by shocks or compressive tides in the stripped gas \citep[e.g.][]{Iglesias12}. The MegaCam  u-band image reveals new ones and provides some details on their structure.   We draw attention on the region referred as SQ-HII on Fig.~\ref{fig:SQ} (bottom-left)  that coincides  with the Northern tip of an H{\sc i} tail, and forms a jet--like stream  with regularly--spaced  giant star complexes.

Several other  filamentary structures are observed in the Eastern region of the Stephan's Quintet. The principle ones are labeled on the top-right panel of Fig.~\ref{fig:SQ}. The tail apparently coming from NGC~7319, known as the inner tail, hosts massive star clusters, that are 150--200 Myr old \citep{Fedotov11} and prominent star-forming condensations, including the TDG candidate, SQ-B \citep{Lisenfeld04}. At the location of a  CO--detected dust lane towards SQ-B, the tail seems to split. One branch,  labeled IT1, extending eastwards, was not known before. It is  diffuse, and lacks star forming regions.
The other  branch, labeled IT2, bifurcates to the North, crossing the other TDG candidate, SQ-tip, and ends in the scattered star-forming knots mentioned above. The difference in gas content (and thus star formation activity) in the two branches remains to be explained.

The so-called outer tail, observed to the South, disappears on its Western side towards the foreground galaxy NGC~7320. The deep CFHT image undoubtedly confirms  that it points to the East towards NGC~7320c, the barred galaxy believed to have crossed the group about 0.5~Gyr ago.
Interestingly, the rather blue colour of this tail (u-r=1.4, see Fig.~\ref{fig:SQ}, bottom-left), testifies the presence of intermediate age stars, but star formation seems to have stopped there: the Eastern tip of the tail is not detected in UV in  GALEX images we collected from the archives. As shown in Fig.~\ref{fig:SQ} (bottom-left), there is no H{\sc i} cloud with a column density larger than  6$\times 10^{19}$ cm$^{-2}$ at this location. 

\subsection{Isolated filamentary structures: cirrus emission}
\label{sec:cirrus}

Fig.~\ref{fig:field} shows a composite image of the whole MegaCam field covering about 1 square degree. Several extended filamentary structures with colours varying between green and yellow (indicating that they are most prominent in the g and r bands) can be seen between SQ in the South, and the spiral galaxy NGC~7331 in the North, in addition to many faint foreground stars, extended ghost halos of bright nearby stars, and background galaxies. 

\begin{figure}
\begin{center}
	\includegraphics[width=0.85\columnwidth]{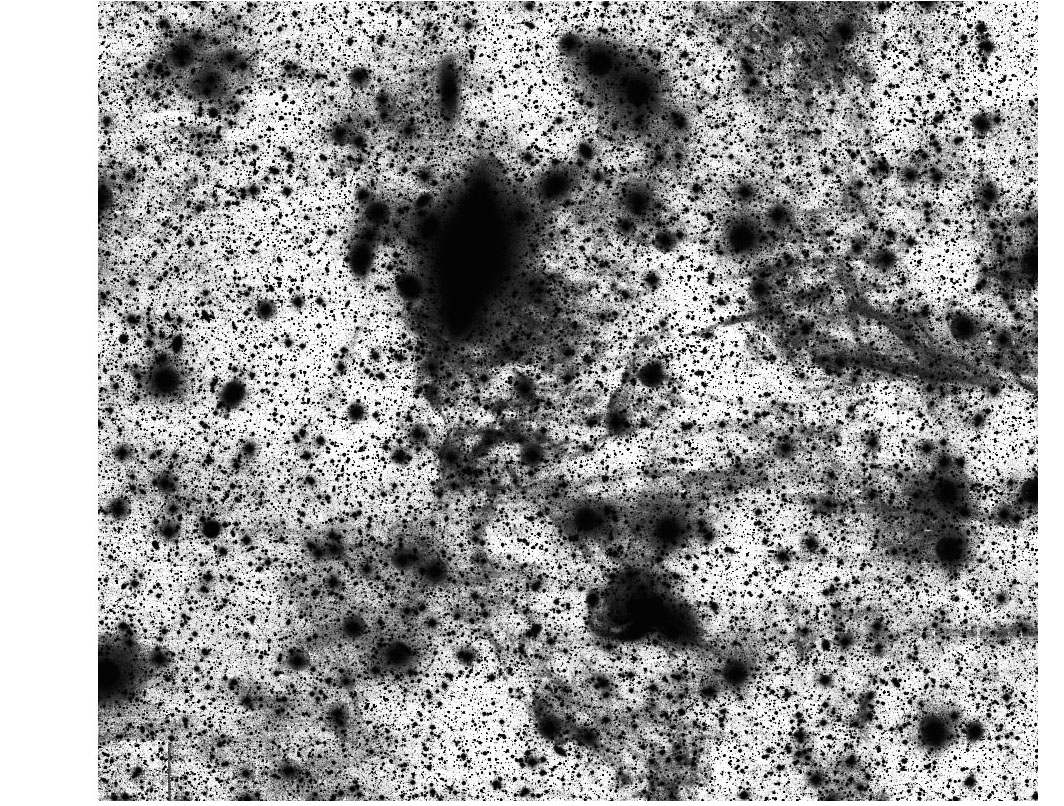}
		\includegraphics[width=0.85\columnwidth]{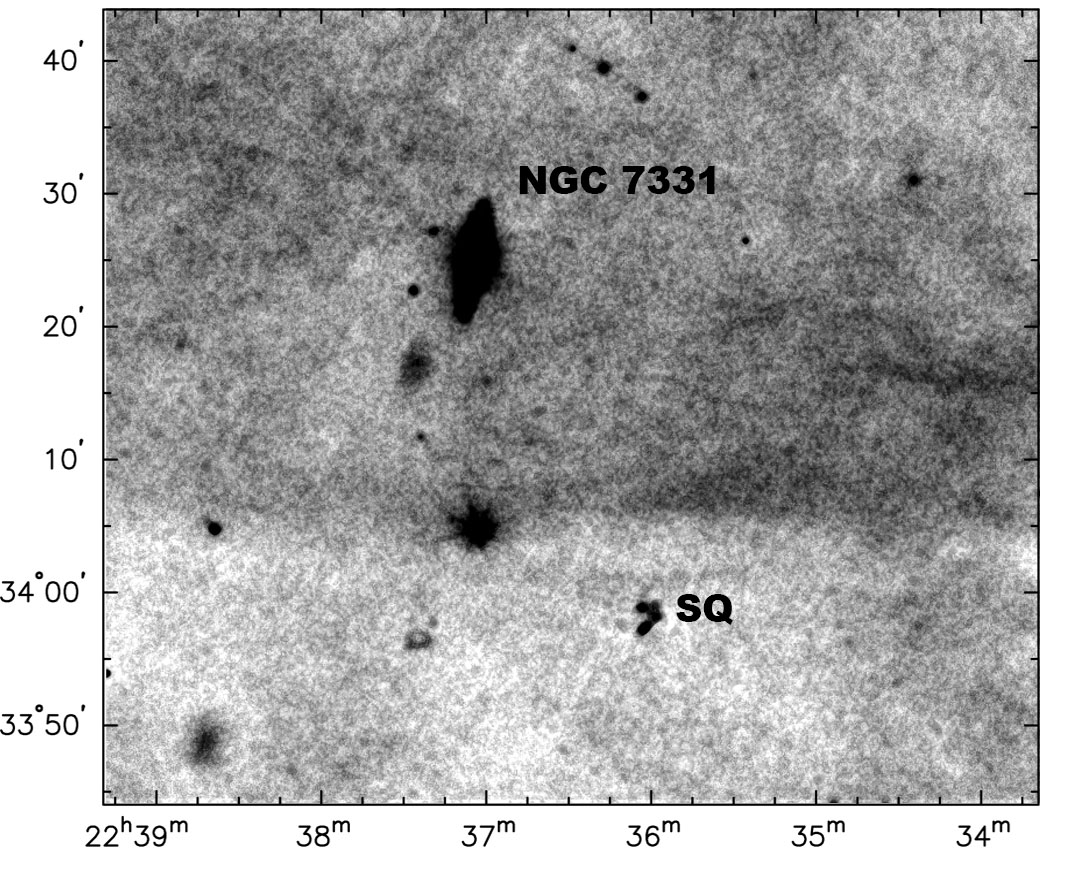}
\end{center}
  \caption{WISE $12\,\mu$m (bottom) and optical CFHT g--band (top)   images of the same field, showing contamination by emission from Galactic cirrus. The RA and DEC coordinates in J2000 are indicated.
  }
  \label{fig:WISE}
\end{figure}

Most of the several isolated narrow filamentary features scattered in the whole field (but more numerous to the North, and globally sharing a East--West orientation) show a clear counterpart in the mid--infrared images (12~$\mu$m, Fig.~\ref{fig:WISE}), which we queried from the WISE archives using the CDS Aladin tool. These structures are thus undoubtedly Galactic cirrus, and not extragalactic tidal structures as initially suggested, among other hypotheses, by \cite{Arp72}. Scattered optical light by dust clouds in the Milky Way hampers extragalactic studies based on deep images, as they mimic stellar streams. They however provide valuable information on the structure of the ISM at small scales, as emphasised by \cite{Miville16}. 

Getting closer to SQ, the plume apparently  dragging on from the diffuse halo of NGC~7317 towards the South-West, and best seen in the r--band, might be a Galactic cirrus as well.
Indeed it has a similar structure and orientation as the other confirmed mid--IR detected  cirrus, though the former is likely  too faint to be detected by  WISE (see Fig.~\ref{fig:WISE}).

\subsection{The spiral galaxy NGC~7331}
NGC~7331 is a highly inclined galaxy with prominent spiral arms. Given its distance (13.5~Mpc, taken from the NED database), it has been argued to form an interacting pair with  NGC~7320 (itself foreground of SQ). However, even at the depth reached here, we do not detect any morphological sign of perturbation in the galactic disk, nor actually the presence of tidally disrupted companions. As mentioned above, the filamentary structures  to the West and South of the spiral are most likely Galactic cirrus and not tidal tails.

\section{Discussion and conclusions}

To date, models of the Stephan's Quintet attribute most of its tidal features to the interactions between NGC~7319, NGC~7320c and NGC 7318~a/b, while the role of the early-type galaxy NGC~7317 is generally ignored due to a lack of observational clues.

We have revisited this system with deep optical multi-band CFHT images which exhibit a number of extended LSB structures, some new, some that were visible in previously published images, but largely ignored. The connection between the outer tail and NGC~7320c is clearly confirmed, proving the active role of this galaxy in the dynamical history of the group.

The reddish halo surrounding SQ, which is most prominent towards NGC~7317, provides indications on the implication of this galaxy in the group's history. The size of the halo, reminiscent of the intracluster light in clusters and fossil groups, could be consistent with a group formation several Gyr ago. A remarkable correspondence between the diffuse component of the X-ray emission (as traced by XMM-Newton) and the diffuse optical light  had been noted by \cite{Trinchieri05}. As shown in Fig.~\ref{fig:SQ} (bottom-right), the spatial matching is best observed for the halo around NGC~7317. If the X-ray emission traces its potential well, this might be another indication that the group is older (and more relaxed) than generally believed. While the peak of the X-ray/optical halo emission matches the position of NGC~7317, the galaxy appears off-centred with respect to the outer isophotes. This disturbed shape could thus be due to a still on-going interaction with NGC~7319.
The presence of a diffuse X-ray emitting hot gas halo suggests that ram pressure could perhaps explain why the tips of the outer tail and Eastern branch of the inner tail are gas poor.

This study shows how deep imaging can, with the proper observing strategy requiring limited telescope time, bring additional constraints on the modelling of interacting systems  by providing  information on the oldest collisional events traced by the old stellar population. More specifically,  all these observed new features call for revised formation scenarios and detailed models of the Stephan's Quintet. 

\section*{Acknowledgements}
Based on observations obtained with MegaPrime / MegaCam, a joint project of CFHT and CEA/DAPNIA, at the Canada-France-Hawaii Telescope (CFHT) which is operated by the National Research Council (NRC) of Canada, the Institut National des Science de l'Univers of the Centre National de la Recherche Scientifique (CNRS) of France, and the University of Hawaii. We warmly thank the CFHT QSO team. 
The data used in this paper were initially made to produce a true color image  shown in the 2018 version of the CFHT calendar produced  in collaboration with Edizioni Scientifiche Coelum (R. Zabotti, G. Anselmi). A crop of it is shown in Fig.~\ref{fig:SQ}.  We thank the  referee for his prompt feedback and useful suggestions.
 FR acknowledges funding from the Knut and Alice Wallemberg Foundation.

\bibliographystyle{mnras}
\bibliography{SQ}

\bsp	
\label{lastpage}
\end{document}